\documentclass[prd,aps,nofootinbib,showpacs]{revtex4}
\usepackage{amsmath}
\usepackage{graphicx}
\usepackage{epsfig}
\def\beq{\begin{equation}}
\def\eeq{\end{equation}}

\def\half{{\textstyle{1\over 2}}}

\def\etal{{\it et al}}

\def\N{{\scriptscriptstyle N}}
\def\mpi{m_\pi}
\def\mo{m}
\def\mn{M_\N}
\def\mno{M}%\def\mno{{\krig M}}
\def\fpi{f_\pi}

\def\ga{g_{\scriptscriptstyle A}}

\def\LEC#1#2{#1^r(#2)}
\begin{document}

\title{Convergence of the chiral expansion for the nucleon mass}
\author{Judith A. McGovern}\email{judith.mcgovern@manchester.ac.uk} 
\author{Michael C. Birse}\email{mike.birse@manchester.ac.uk}
\vskip 20pt
\affiliation{Theoretical Physics Group, School of Physics and Astronomy\\
The University of Manchester, Manchester, M13 9PL, U.K.}
\pacs{12.38.Gc, 12.39.Fe, 14.20.Dh}

\begin{abstract}
A number of papers recently have used fourth-order chiral perturbation theory to extrapolate lattice data for the nucleon mass; the process seems surprisingly successful even for large pion masses.  This paper shows that the inclusion of the fifth-order terms spoils the agreement.
\end{abstract}
\maketitle

Over the last few years there has been an explosion of activity in the field of chiral extrapolations of lattice QCD data.  However for many quantities of interest unquenched calculations have typically only been performed at relatively high quark masses, so that the pion mass is 500~MeV or more.  There are real questions about the convergence of chiral expansions in this region.

One quantity for which the chiral expansion has seemed to work surprisingly well is the nucleon mass. Various groups have looked at the $O(p^4)$ (technically N$^3$LO) expansion of the nucleon mass in heavy baryon chiral perturbation theory (HB$\chi$PT) and found good agreement with data up to pion masses of 800~MeV or more.  Of course the results are quite sensitive to the input parameters which include some rather poorly known low-energy constants, but when these are used as fit parameters the results agree well with other determinations.

Some years ago the $O(p^5)$ contribution in HB$\chi$PT was calculated \cite{MB99}.  At this order two-loop graphs enter, but it turns out that there are no irreducible two-loop contributions and all the reducible graphs can be absorbed into the physical pion mass and the pion-nucleon coupling constant in the third-order contribution.  Only a relativistic ($1/\mn$) correction is left, and it is tiny: just 0.3\%  
of the third-order term for $\mpi=139$~MeV.

The details of this calculation have not so far been used by any group exploring the chiral expansion of the nucleon mass. Refs.~\cite{PHW04,A-K04,PMWHW06} include the relativistic correction (which can also be obtained from a fully relativistic one-loop calculation), while Ref.~\cite{Mei06} uses its smallness to support the view that $O(p^5)$ contributions will not spoil the convergence. However this is not correct.  So long as one is interested in comparing the predictions of  $\chi$PT to experiment in the real world, it is both legitimate and necessary recast the results of calculations, obtained initially in terms of bare or chiral limit parameters, in terms of their physical values.  But if one wishes to explore the variation with quark mass in lattice calculations, the expressions must be left in terms of bare parameters.  (If these are not well-known they may in practice be given the same numerical value as their real-world equivalents, but that is a different matter.)

Ref.~\cite{MB99} was not written with lattice extrapolations in mind, and so the appropriate expression was not given explicitly.  Furthermore the calculation was performed using the Lagrangian and notation of Ref.~\cite{FMS98}.  This non-minimal Lagrangian contains more parameters (31) than there are independent LECs (23), so the extra ones are taken to vanish at a particular renormalisation scale -- the physical pion mass.  This is an unfortunate choice if the aim is to make explicit the running with (lattice) pion mass.  The aim of the current paper is to present the $O(p^5)$ result in an appropriate form for comparison with lattice results, and to explore its consequences for the convergence of the chiral expansion. We should remark that Beane has also opined that fifth-order corrections could be significant, using a rather different argument \cite{beane04}.

The nucleon mass in the chiral limit is denoted by $\mno$. The next four terms in the chiral expansion for non-vanishing quark mass, and hence bare pion mass $\mo$, are as follows:
\begin{eqnarray}
\delta \mn^{(2+3)}&\!=\!&-4 c_1 \mo^2 - {3g^2\over 32\pi f^2}\mo^3\nonumber\\
\delta \mn^{(4)}&\!=\!&\mo^4\left(- e(\lambda)+
\frac{3}{128\pi^2 \fpi^2}\left(c_2-\frac{2\ga^2}{\mno}\right)
%\right.\nonumber\\ &&\left.
-\frac{3}{32 \pi^2 \fpi^2 }\left(\frac{\ga^2}{\mno} - 8 c_1+c_2+4c_3\right)\ln
{\frac{\mo}{\lambda}}\right)\nonumber\\
\delta \mn^{(5)}&\!=\!&{3g^2\mo^5\over 32\pi f^2}\left({2\LEC{l_4}\lambda-3\LEC{l_3}\lambda\over f^2}-
{4(2 \LEC{d_{16}}\lambda- d_{18})\over g}+16\LEC{d_{28}}\lambda+{16g^2-3\over 32\pi^2f^2}\ln{m\over \lambda} 
+ %\right.\nonumber\\&&\hspace*{3cm}+\left.
{g^2\over 8\pi^2f^2}+{1\over 8\mno^2}\right).
\end{eqnarray}

The parameters $g$, $f$ and $\mo$ are the leading terms in the chiral expansions of $\ga$, $\fpi $ and $\mpi$; $m^2$ is proportional to the quark mass. The physical values of these parameters are $\ga^{\rm phys}=1.267$, $\fpi^{\rm phys}=92.4$~MeV and $\mpi^{\rm phys}=139.6$~MeV; in what follows the bare parameters $g$ and $f$ will be set to the physical values  as the chiral-limit values are not well known. The second-order LECs $c_i$ are more-or-less well determined from pion-nucleon and nucleon-nucleon scattering. Since this paper follows most closely the methods of Refs.~\cite{A-K04,PMWHW06} we use the same values as there: $c_1$ is generally fitted but has been determined to be in the range $-1.1$ to $-0.4$~GeV${}^{-1}$; $c_2=3.2$~GeV${}^{-1}$ and $c_3=-3.4$~GeV${}^{-1}$; the exact values of the latter two do not matter because they only occur along-side higher-order LECs which will be fitted.  

The finite third-order LEC ${d}_{18}$ is fitted to the Goldberger-Trieman discrepancy, and using $g_{\pi\N\N}=13.18$ gives ${d}_{18}=-0.78\pm0.27$~GeV$^{-2}$ \cite{FMS98}.  With this as input, the value of $\LEC{d_{16}}{\mpi^{\rm phys}}$ is extracted from the reaction $\pi{\rm N}\to\pi\pi{\rm N}$ in Ref.~\cite{FBM00}; however the same authors subsequently use a substantially revised value, and Ref.~\cite{beane04} quotes their 1 $\sigma$ range to be $-2.61$ to $-0.17$~GeV$^{-2}$.

The notation for the mesonic LECs is  
$\LEC{l_i}\lambda=(\gamma_i/32\pi^2)\,(\overline{l}_i+2\ln{\mpi^{\rm phys}/\lambda})$, with $\overline{l}_3=2.9\pm2.5$,
$\overline{l}_4=4.4\pm0.3$, $\gamma_3=-\half$ and $\gamma_4=2$ \cite{GL84}.

By $e(\lambda)$ we denote a combination of LECs: $e(\lambda)=16\LEC{e_{38}}\lambda+2\LEC{e_{115}}\lambda+2\LEC{e_{116}}\lambda-32 c_1 \LEC{d_{28}}\lambda$, where $d_i$ and $e_i$ are third- and fourth-order LECs respectively.

The fourth-order calculation was first given in Ref.~\cite{SMF98} (where fourth-order LECs were denoted $b_i$).  The fifth-order calculation is from Ref.~\cite{MB99} with 
$\LEC{d_{28}}\lambda$ restored; it was omitted there because it vanishes at $\lambda=\mpi^{\rm phys}$. An unconventional notation for the mesonic LECs was also used in that paper.

If lattice results presented the variation of $\mn$ with the quark mass, the expression above would be the relevant one.  Since, however, the usual presentations plot $\mn$ against $\mpi$, both being lattice measurements, the correct expression uses the NLO---$O(p^4)$---running of the pion mass to give 
\begin{eqnarray}
	\delta \mn^{(2+4)}
	&\!=\!& -4c_1 \mpi^2
+%\nonumber\\&\!+\!&  
\mpi^4\left[- e'(\lambda)+
\frac{3}{128\pi^2 \fpi^2}\left(c_2-\frac{2\ga^2}{\mno}\right)
%\right.\nonumber\\ &&\left. 
-\frac{1}{32 \pi^2 \fpi^2 }\left(\frac{3\ga^2}{\mno} - 32 c_1+3c_2+12c_3\right)\ln{\frac{\mpi}{\lambda}}\right]\nonumber\\
\delta \mn^{(3+5)}&\!=\!&
{3g^2\mpi^3\over 32\pi f^2}\left[-1+(\mpi)^2\left({2\LEC{l_4}\lambda\over f^2}-
{4(2 \LEC{d_{16}}\lambda- d_{18})\over g}+16\LEC{d_{28}}\lambda+{g^2\over 2\pi^2f^2}\ln{\mpi\over \lambda}
+%\right.\right.\nonumber\\&&\hspace*{3cm}+\left.\left.
{g^2\over 8\pi^2f^2}+{1\over 8\mno^2}\right)\right],
\label{fullrun}
\end{eqnarray}
with $e'(\lambda)=e(\lambda)+8 c_1 \LEC{l_3}\lambda/f^2$.
At fourth order it make very little difference which expression one uses.

The LECs that enter  can, in principle, all be extracted from $\pi\pi$, $\pi {\rm N}$ and NN scattering data and the physical nucleon mass.  As indicated above, most already have been, with greater or lesser uncertainties.  However the fourth-order LEC $e'(\lambda)$ is very poorly constrained and is generally fitted to lattice data.  Bernard \etal\  recognised that with one free parameter the fourth-order expression could fit lattice data out to $\mpi$ of about 600~MeV \cite{BHM04}; Procura \etal\ showed that the best fit to the lattice data up to 600~MeV was actually a superb fit right up to much higher masses (though without much predictive power when one varied the fit parameters within acceptable limits of $\chi^2$) \cite{PMWHW06}.  From such results it has been concluded that the chiral expansion may be valid much further than naive estimates might suggest.  However until now the full fifth-order terms have not been included.

\begin{figure}[htb]
\begin{center}
\epsfig{file=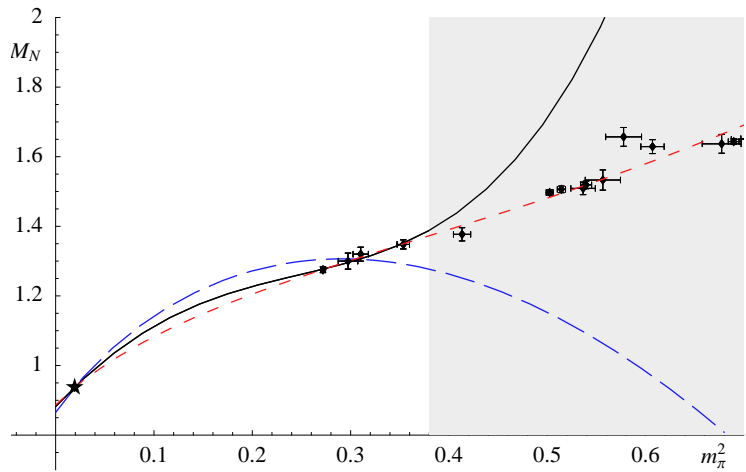,width=10cm}
\end{center}
\caption{Best fits to the lattice data (constraint to pass through the physical point) for $\mn$ (in GeV) versus $\mpi^2$ (in GeV$^2$) below $\mpi=600$~MeV (unshaded region) at third- (blue, long dashes), fourth- (red, short dashes) and fifth order (black, solid).  See text for the sources of the data. (Note that the apparent agreement of the fourth- and fifth-order curves at low $\mpi^2$ masks very different fitted values of the LEC $e'$.)}
\label{massfig1}
\end{figure}

In Fig.~\ref{massfig1} we show the analogues of the fit of Procura \etal\ \cite{PMWHW06} at third, fourth and fifth order (Eq.~\ref{fullrun}). The lattice data is from the following collaborations: UKQCD and QCDSF \cite{A-K04}, CP-PACS \cite{A-K02,A-K04}, JLQCD \cite{Aoki03,A-K04} and Orth, Lippert and Schilling \cite{OLS05}; the selection of points at large enough lattice volume is made by Musch in Ref.~\cite{Mus06}; see also Ref.~\cite{PMWHW06}. 

Like Procura \etal, at fourth order we fit $\mno$ and $e'(\mpi^{\rm phys})$ to the lattice data below 600~MeV, choosing $c_1$ so that the physical $\mn$ is reproduced.  At third order in HB$\chi$PT only $\mno$ is free.  At fifth order we choose to fit $\LEC{d_{16}}{\mpi^{\rm phys}}$ since it is poorly determined otherwise.  However treating $\mno$ (or equivalently $c_1$) as a fit parameter gives the unacceptable solution $\mno>\mn$, so instead we fix $c_1$ at the value emerging from the fourth-order fit.  

We see that at third order an acceptable fit cannot be obtained: $\chi^2=31.6$ for 3 d.o.f.
However the value $c_1=- 1.14$~GeV$^{-1}$ ($\mno=865$~MeV) is quite reasonable.  At fourth order the fit is extremely good, with $\chi^2=0.20$ for 2 d.o.f..  Also, the value $c_1=-0.84$~GeV$^{-1}$ is within the range obtained by other means and $e'(\mpi^{\rm phys})=-0.52$~GeV$^{-3}$ is natural in size.\footnote{This is equivalent to $e'(\hbox{1 GeV})=-3.67$GeV$^{-3}$. The slight deviation from the values found by Procura \etal\  is due to the fact that they did not include the corrections for the running of $\mpi$ and also to their inclusion of the ``naive" fifth-order term.}  At fifth order the fit is equally acceptable: $\chi^2=0.54$ for 2 d.o.f. However the fitted values are well out of line with what one would expect: $\LEC{d_{16}}{\mpi^{\rm phys}}=4.654$~GeV$^{-2}$ and $e'(\mpi^{\rm phys})=-30.5$~GeV$^{-3}$.  Furthermore in sharp contrast to the fourth-order result, the fifth-order curve, though good for the points it was fitted to, breaks down at higher values of $\mpi$.  Allowing the fit to vary within an acceptable range of $\chi^2$ does not change this behaviour.  Whereas the fourth-order fit results in distinctly small coefficients,
$\mn=0.887+ 3.35\mpi^2 -5.61\mpi^3+ 3.96\mpi^4$ (in GeV), the fifth-order fit does not:
$\mn=0.885+ 3.20\mpi^2 -5.61\mpi^3+ 33.9\mpi^4 -108.5\mpi^5$. (The logs are evaluated at $\mpi=0.2$~MeV but do not change the qualitative picture). In the latter the large size of the fifth-order term is not an artifact of the unrealistic value of $\LEC{d_{16}}{\mpi^{\rm phys}}$; with this set to zero the coefficient is still $56$.  (Interestingly though, the order of magnitude is not so very different from that found ($\approx -40$~GeV$^{-4}$) in most of the finite-range regularised expressions used to extrapolate over a very large range of pion masses by Leinweber \etal\ \cite{LTY05}.

\begin{figure}[htb]
\begin{center}
\epsfig{file=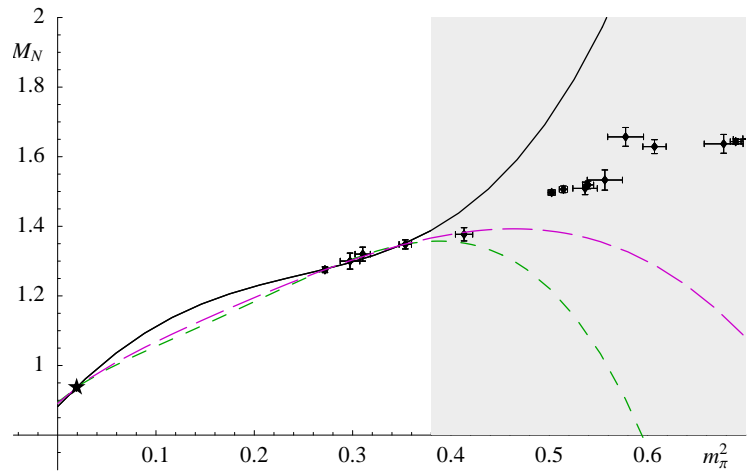,width=10cm}
\end{center}
\caption{Best fits to the data below $\mpi=600$~MeV at fifth order (black, solid), without allowing $\ga$ to run (magenta, long dashes), and with a 6th-order term designed to fit the lattice running of $\ga$ (green, short dashes). See text for more details.}
\label{massfig2}
\end{figure}

As already mentioned, the bulk of the fifth-order term comes from the running with $\mpi$ of the pion-nucleon coupling constant, and in particular of $\ga$.  But it is well known that NLO $\chi$PT does not do a good job of reproducing lattice data on $\ga$; the best fit to date has included an explicit $\Delta$ \cite{A-K06}, which is absent here.  As an experiment, we added an $\mpi^6$ term to the expression for the nucleon mass as if it came from the $\mpi^3$ term in $\ga$, with a coefficient chosen to give a reasonable fit to the lattice data presented in Ref.~\cite{A-K06}.\footnote{This is achieved with $\LEC{d_{16}}{\mpi^{\rm phys}}=1.24$~GeV$^{-2}$ and (in $\ga$ only) $c_3-2c_4=0.62$~GeV$^{-1}$. The third-order expansion of $\ga$ was first given in Ref.~\cite{BKKM92}.} Alternatively we simply dropped all terms in the running of $\mn$ originating from the running of $\ga$ (since the latter is apparently quite weak). The results are shown in Fig.~\ref{massfig2}. Although the resulting curves are quite different from the standard one, they cannot be said to be much more successful at extrapolating beyond the fit region, nor are the corresponding extracted LECs notably more acceptable. 

Ultimately, we do not consider the results of this paper to be a surprise.  What was surprising was the original finding that HBCPT appeared to work not only out to 600~MeV but even beyond. Our results imply that lattice data will be required at substantially lower pion masses before chiral extrapolations will be reliable.

\begin{acknowledgments}
J. McG would like to thank Prof.~A.~Thomas for prompting her to look again at the fifth-order calculation with lattice extrapolation in mind.
\end{acknowledgments}

\end{document}